\documentclass{optica-article}

\journal{opticajournal} 

\articletype{Research Article}

\usepackage{lineno}
\usepackage{amsmath}
\usepackage{gensymb}
\usepackage[section]{placeins}

\begin{document}

\title{Three-dimensional reconstruction of THz near-fields from a LiNbO$_3$ optical rectification source}

\author{Annika E. Gabriel,\authormark{1,2*} Mohamed A.K. Othman,\authormark{1} Patrick L. Kramer,\authormark{1} Harumy Miura,\authormark{1} Matthias C. Hoffmann,\authormark{1} and Emilio A. Nanni\authormark{1}}

\address{\authormark{1}SLAC National Accelerator Laboratory, Stanford University, Menlo Park, California 94025, USA\\
\authormark{2}Department of Physics, University of California Santa Cruz, Santa Cruz, CA, USA.}

\email{\authormark{*}angabrie@slac.stanford.edu} 

\begin{abstract*} 
Terahertz (THz) generation by optical rectification in LiNbO$_3$ (LN) is a widely used technique for generating intense THz radiation. The spatiotemporal characterization of THz pulses from these sources is currently limited to far-field methods. While simulations of tilted pulse front THz generation have been published, little work has been done to measure the near-field properties of the THz source. A better understanding of the THz near-field properties will improve optimization of THz generation efficiency, transport, and coupling. We demonstrate a technique for quantitative spatiotemporal characterization of single-cycle strong-field THz pulses with 2-D near-field electro-optic imaging. We have reconstructed the full temporal 3D THz near-field and shown how the phase front can be tailored by controlling the incident pump pulse.

\end{abstract*}

\section{Introduction}
One of the most popular techniques for high-field THz generation is optical rectification in LN using the tilted pulse front method \cite{Leitenstorfer_2023,Dhillon_2017, Hebling:08Rev}. It is routinely used to generate THz pulses with energies in the micro-joule range and field strengths above 1 MV/cm \cite{Dhillon_2017}.  The development of this reliable tabletop method has led to applications in strong field spectroscopy, time-domain spectroscopy, and nondestructive testing \cite{Hebling:02,Yeh,Hebling:08,Hirori, Koch:23, Teo:15, Baxter:11, Nsengiyumva:23}. Harnessing THz radiation for all of these applications requires coupling THz beams out of the LN crystal and maximizing conversion efficiency \cite{Leitenstorfer_2023}. In addition, THz radiation from tilted pulse front sources is being used in promising new techniques for electron beam timing diagnostics, acceleration, bunch compression, and ultrafast photoguns \cite{Othman:23, Hibberd:2019okk, L.J.R.Nix:24, Nanni, Snively, Li, Ying:24}. In these techniques, the THz pulse interacts with a charged particle beam requiring phase-matching between the particles' velocity and the THz field. This requires optics that both transport the appropriate spectrum and provide the correct phase front. A major challenge for these THz applications is power losses during the transport of single-cycle THz pulses using free-space optics, and when coupling into vacuum chambers or waveguide structures \cite{Hibberd:2019okk,Xu:2020egd}. Mapping of the near-field enables the design of specialized transport optics that could significantly improve coupling efficiency, allowing higher field strengths and a broader frequency range at the focus. In addition, imaging of the THz near-field yields insight into the factors affecting THz conversion efficiency by illuminating the direct effect of different tilted pulse front parameters on the generated THz near-field and its spectral distribution. 

We have developed a technique for detailed quantitative measurement of the THz near-fields with sub-wavelength resolution that reveals the effects caused by varying the optical rectification setup. To the best of our knowledge, no other measurements of the THz near-field for a tilted pulse front source have been conducted with this level of precision. Our imaging technique utilizes electro-optic sampling to produce 3D spatiotemporal data of the THz near-field at the output face of the LN prism. From this data it is possible to calculate the THz field strength and map the full evolution of the field.  In most THz experiments a series of off axis parabolic mirrors are used to transport and focus the THz beam from the source onto an electro-optic crystal, such as gallium phosphide (GaP) or zinc telluride (ZnTe)\cite{Wu1995}. The THz field then induces a change in refractive index within the GaP which produces a change in the polarization state of a probe femtosecond laser pulse \cite{Adam}. Measuring the degree of ellipticity of the probe laser pulse yields the temporal structure and relative field strength of the THz pulse at that location \cite{Planken:01}. In standard EOS the signal either averages over the THz wavefront if the probe beam is large compared to the THz spot size, or it only samples a subsection of the field if the probe beam is small compared to the THz spot size on the GaP. Electro-optic imaging of the THz beam at the focus of a parabolic mirror has been demonstrated in Ref. \cite{Wu1996}.

In our approach, we utilize an optimized probe beam with a large spot size incident on a GaP crystal placed within the near-field at the THz emission surface of the LN crystal. A coating on the back of the GaP allows the probe beam to be reflected and captured on a CCD camera, effectively imaging the entire THz emission area in one shot. This technique provides complete quantitative spatiotemporal characterization of the THz near-fields and quick acquisition of the data needed for a thorough reconstruction. It also yields significant improvements in the overall characterization of the THz near-field when compared to previous techniques \cite{Annika22,Gabriel:22}, with 3 orders of magnitude higher spatial resolution and a reduction in acquisition time from minutes to less than a second for a single image.  The robust characterization of the near-field could inform designs of novel structures for use in THz time domain spectroscopy, THz particle acceleration, and ultrafast particle beam manipulation. In addition, these results can illuminate a deeper understanding of the THz generation mechanism and the impact of parasitic nonlinear effects such as self-phase modulation \cite{Ravi:15}.

\section{Methods}
\begin{figure}[htbp]
\centering\includegraphics[scale=0.35]{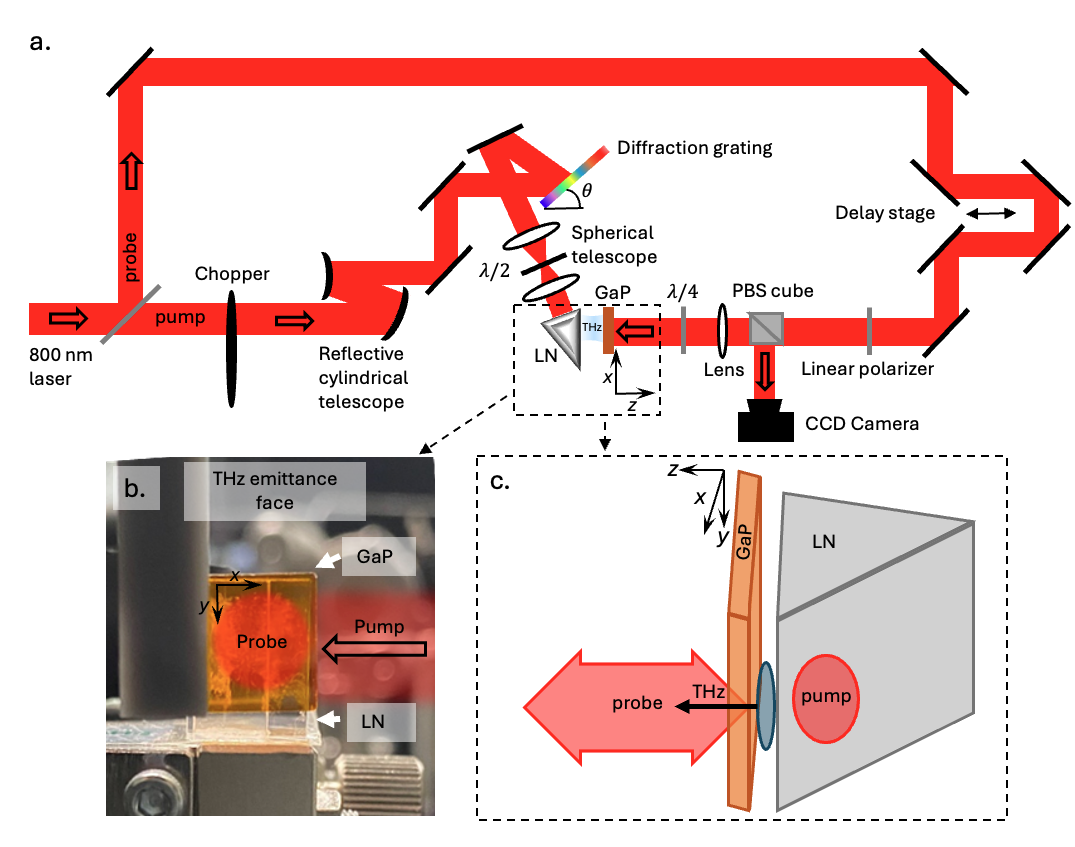}
\caption{The full experimental setup includes (a) THz generation with tilted pulse fronts, chopper for background subtraction, delay stage to vary time of arrival of EOS probe, and detection using a CCD camera. Additionally, an image of the EOS setup is shown in (b) with illustrations of the pump and probe beam directions and locations. The LN and GaP crystal from the angle of incidence of the 800 nm pump is shown in (c).}
\label{experiment_diagram}
\end{figure}

In our experiments we characterized the THz near-field of a tilted pulse front source at the exit face of the LN crystal. THz generation was conducted using optical rectification of Ti:Sapphire laser pulses with $\sim2$~mJ per pulse, $\sim 50$ fs pulse duration, 800~nm central wavelength, operating at 1~kHz. A diffraction grating with 2000~lines/mm was used to produce the pulse front tilt and a telescope, composed of 100 mm and 50 mm focal length spherical lenses, imaged the tilted pulse front onto the LN, further increasing the tilt angle (Fig. \ref{experiment_diagram}a). Initial optimization of the THz conversion process was done using a pyroelectric detector at the output face of the LN prism. A chopper was installed in the pump beamline so that the imaging camera could be triggered to capture images of the probe beam with the pump beam blocked (THz OFF) and not blocked (THz ON) for background subtraction. 

Our imaging technique measures the THz near-field using EOS in a <110> cut 10~mm × 10~mm × 0.5~mm GaP crystal (Fig. \ref{experiment_diagram} b, c). The GaP crystal was coated with a highly reflective coating on one side and an anti-reflective coating on the other, both at 800~nm wavelength. The side of the GaP crystal with the highly reflective coating was placed facing towards and parallel to the exit plane of the LN with a separation distance of 0.6~mm. This separation distance is well within the near-field of this source which is $\sim$6.7~mm from the exit face of the LN. The range of the near-field for this source was calculated from the well known near-field definition $2D^2/\lambda$ where $D$ is the aperture of the source ($\sim$2~mm) and $\lambda$ is the wavelength of the central frequency ($\sim$0.5~THz, $\lambda=$0.6~mm). The incoming probe beam polarization was controlled using a linear polarizer, half-wave plate, polarizing beam-splitter cube, and a quarter-wave plate. The probe beam was then aimed onto the GaP crystal so that it encompassed the entire THz emission area. After reflection off the back face of the GaP crystal the vertical component of the tilted laser pulse is imaged using a lens and 1:1 magnification onto a CCD camera. The spatial resolution of the camera was $4.5 \pm 0.1$ and $5.5 \pm 0.2$ $\mu$m/pixel in the vertical (y) and horizontal (x) direction respectively.

\begin{figure}[htbp] 
\centering\includegraphics[scale=0.3]{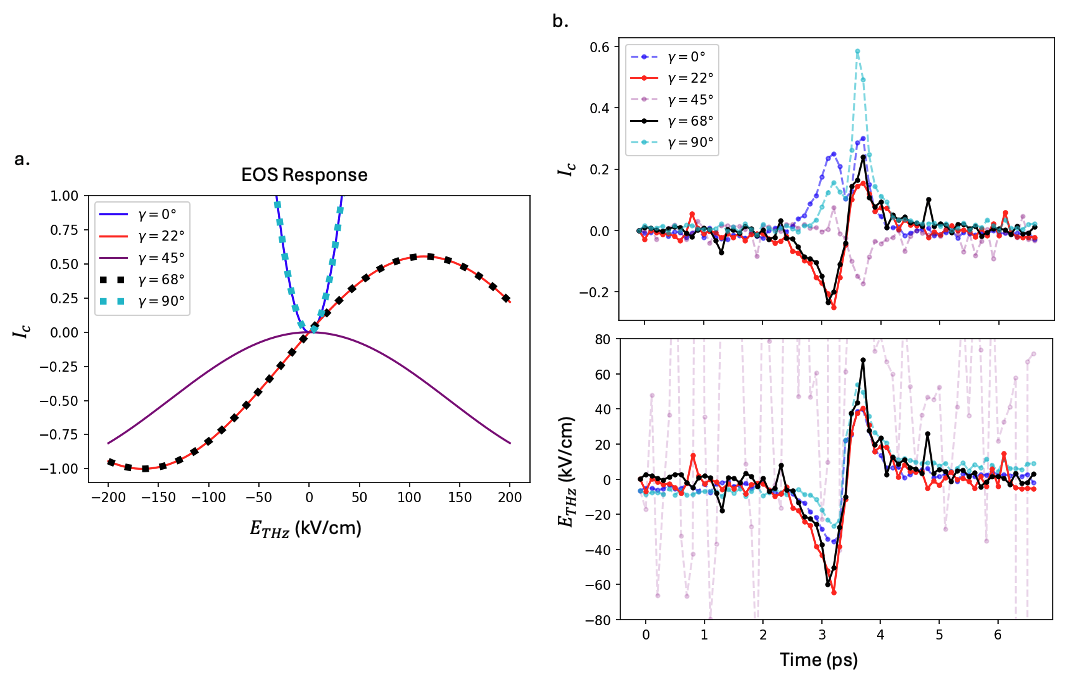}
\caption{The calculated intensity on the camera ($I_c$) after background subtraction as a function of THz field strength for different quarter waveplate angles ($\gamma$) is shown in (a). The top plot in (b) shows THz time traces taken at the location of peak field for each quarter waveplate angle, and the bottom plot shows these same time traces when scaled by the theory in (a). Based on this data it can be seen that $\gamma=22\degree,\ 68\degree$ are in the linear EOS regime and $\gamma=0\degree,\ 45\degree,\ 90\degree$ are nonlinear.}
\label{EOS_theory}
\end{figure}

The probe polarization necessary to produce a linear EOS response was calculated (Fig. \ref{EOS_theory}a) using the standard Jones matrix technique and known THz EOS theory \cite{Planken:01}, for details see our supplementary information. The calculation was done with respect to $\gamma$, the angle between the quarter waveplate fast axis and the horizontal axis parallel to the optical table. We calculated the relation between the normalized intensity measured on the camera $I_c$ and the actual THz field ($E_{THz}$) response for a series of angles from $\gamma=0\degree$ (fast axis parallel to the table) to $\gamma=90\degree$ (fast axis perpendicular to the table). The normalized intensity $I_c = I_{B,\gamma} - I_{THz} / I_{B,\gamma}$ is calculated from the background intensity, $I_B$ (THz off), for each value of $\gamma$, and $I_{THz}$ (THz on). It was only necessary to calculate the response from $0\degree - 90\degree$ due to the inherent periodicity of the EOS system.  Based on these results it can be seen that for quarter waveplate angles $\gamma = 22\degree,\ 68\degree$ the EOS response is in the linear regime for the produced THz fields. In contrast, $\gamma = 0\degree,\ 45\degree,\ \textrm{and}\ 90\degree$ are in the nonlinear regime and produce unipolar responses in either the negative or positive direction. This calculated relation between measured intensity on the camera ($I_c$) and actual THz field strength ($E_{THz}$) can be determined for the measured THz time traces at various $\gamma$ values to produce THz field values (Fig. \ref{EOS_theory} b), see supplement for detailed derivation. These results clearly show a typical THz time trace when $\gamma$ is in the linear range ($22\degree, 68\degree$). It also converts the unipolar responses of $\gamma=0\degree,90\degree$ to bipolar traces that show excellent agreement with the linear response. In the $\gamma=45\degree$ case, the quarter wave plate angle allowed more scattered light to pass into the measurement due to reflections from the pump beam. This noise was then amplified when converted to THz field, showing that any measurements taken in this region would be highly inaccurate and lead to the signal being overwhelmed by noise.

\begin{figure}[htbp] 
\centering\includegraphics[scale=0.5]{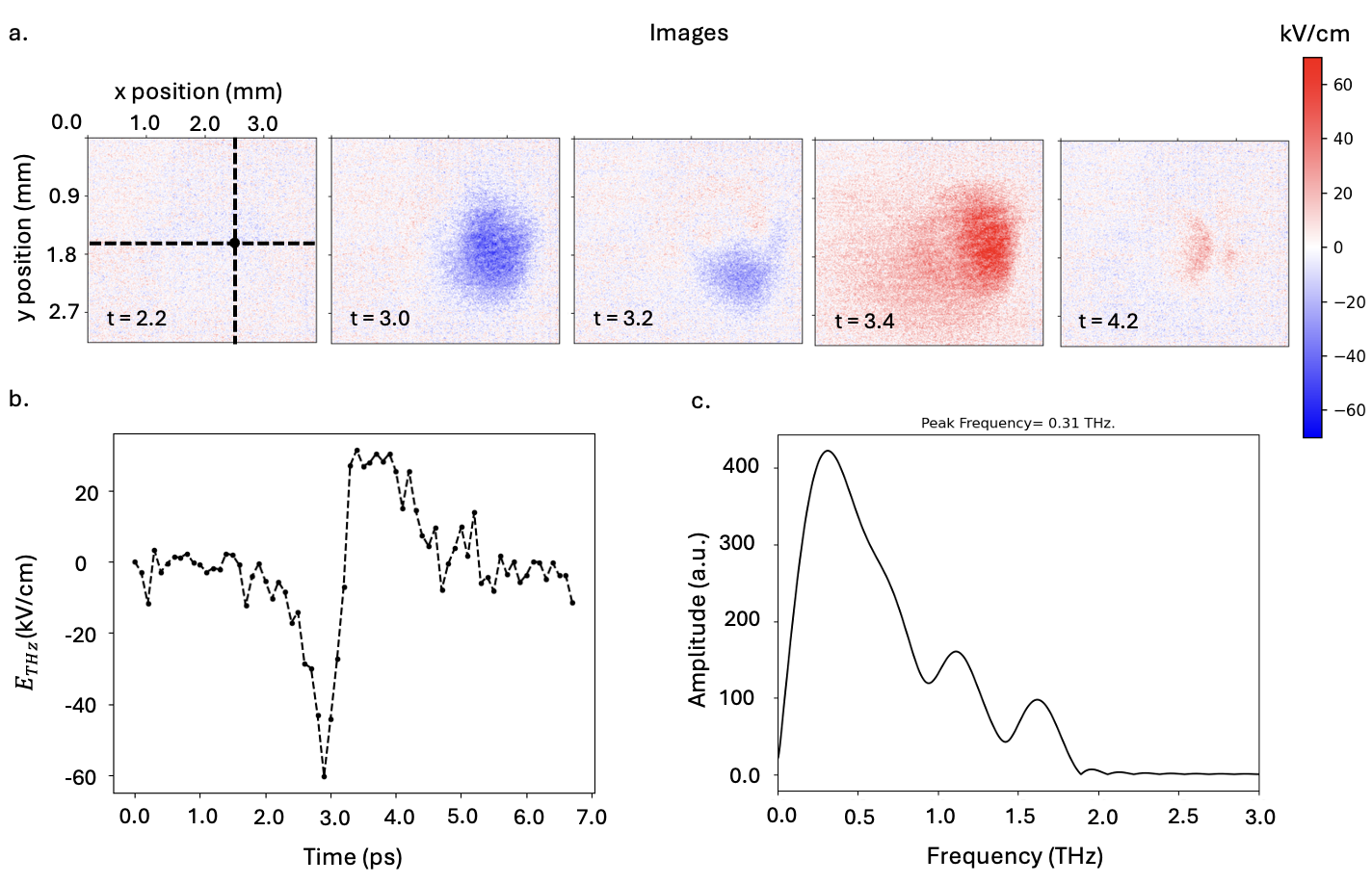}
\caption{Example data of the ideal configuration showing images after background subtraction, normalization, and filtering in (a), time trace in (b), and frequency spectrum in (c). The time trace in (b) was extracted at the location of absolute maximum THz field, marked by the dashed line cross in the first image of (a). The frequency spectrum in (c) was obtained by Fourier transforming the time trace in (b). This data yielded a minimum and maximum THz field strength of -64.0 kV/cm and 56.1 kV/cm respectively. The calculated peak frequency was 0.31 THz centered around mean frequency 0.67 THz.}
\label{Ideal_THz}
\end{figure}

Our imaging technique allows the reconstruction of the full THz pulse (Fig. \ref{Ideal_THz}) with scans typically taking on the order of minutes to complete depending on the desired temporal resolution. The quarter waveplate angle was set at $\gamma=22\degree$, allowing calculation of the THz field strength based on the measured intensity values on the camera with high accuracy. Scans of the full THz field temporal evolution with 10~fs time resolution can be completed in under an hour, and even more precise timing resolution was possible up to the precision of the time delay translation stage. Figure \ref{Ideal_THz} shows images of the THz field at various time delays when our measurement technique was used to optimize the generation setup to produce a phase front that was parallel to the surface of the LN. A temporal trace at the location of absolute maximum THz field (Fig.~\ref{Ideal_THz}b) can be extracted by integrating over a fixed area of pixels, marked by the dashed line in Fig. \ref{Ideal_THz}a and the corresponding frequency spectrum can be obtained by a Fourier transform (Fig.~\ref{Ideal_THz}c). The Gaussian spatial field profile, temporal behavior, field strength, and frequency spectrum are all within the expected range for tilted pulse front THz generation predicted from standard far-field measurements \cite{Adam}. These results confirm the validity of the method and show we are operating in the linear EOS regime.

\FloatBarrier

\FloatBarrier

\FloatBarrier

\section{Results}
The expected spatial profile of the THz near-field under optimum conditions is the input pump beam profile projected on a 63° angle. In our experiment the pump beam was shaped with a reflective cylindrical telescope before the diffraction grating so that the THz beam is expected to be Gaussian in the spatial dimensions and emitted orthogonally to the 2D LN crystal face. The ideal generation settings were determined by optimizing to produce a pulse with these properties using our imaging method. Figure \ref{diffraction_tilt_data} shows images of the THz near-field for  the optimum grating tilt angle at several time delays. The spatial profile is Gaussian and the peak field strength calculated from this data was -64.0 kV/cm with a peak frequency of 0.31 THz centered around a mean frequency of 0.67 THz (Fig.\ref{Ideal_THz}).

\begin{figure}[htbp] 
\centering\includegraphics[scale=0.3]{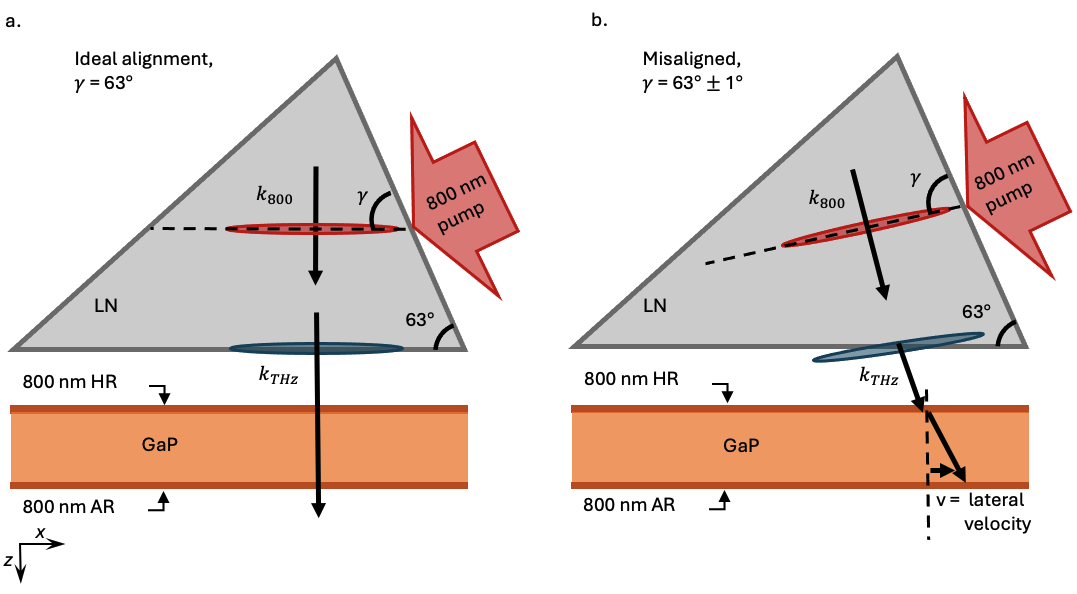} 
\caption{Diagram illustrating the effect of tilting the diffraction grating angle on the generated THz field where (a) is ideal THz generation and (b) is slight misalignment.}
\label{diffraction_tilt_diagram}
\end{figure}

\begin{figure}[htbp] 
\centering\includegraphics[scale=0.35]{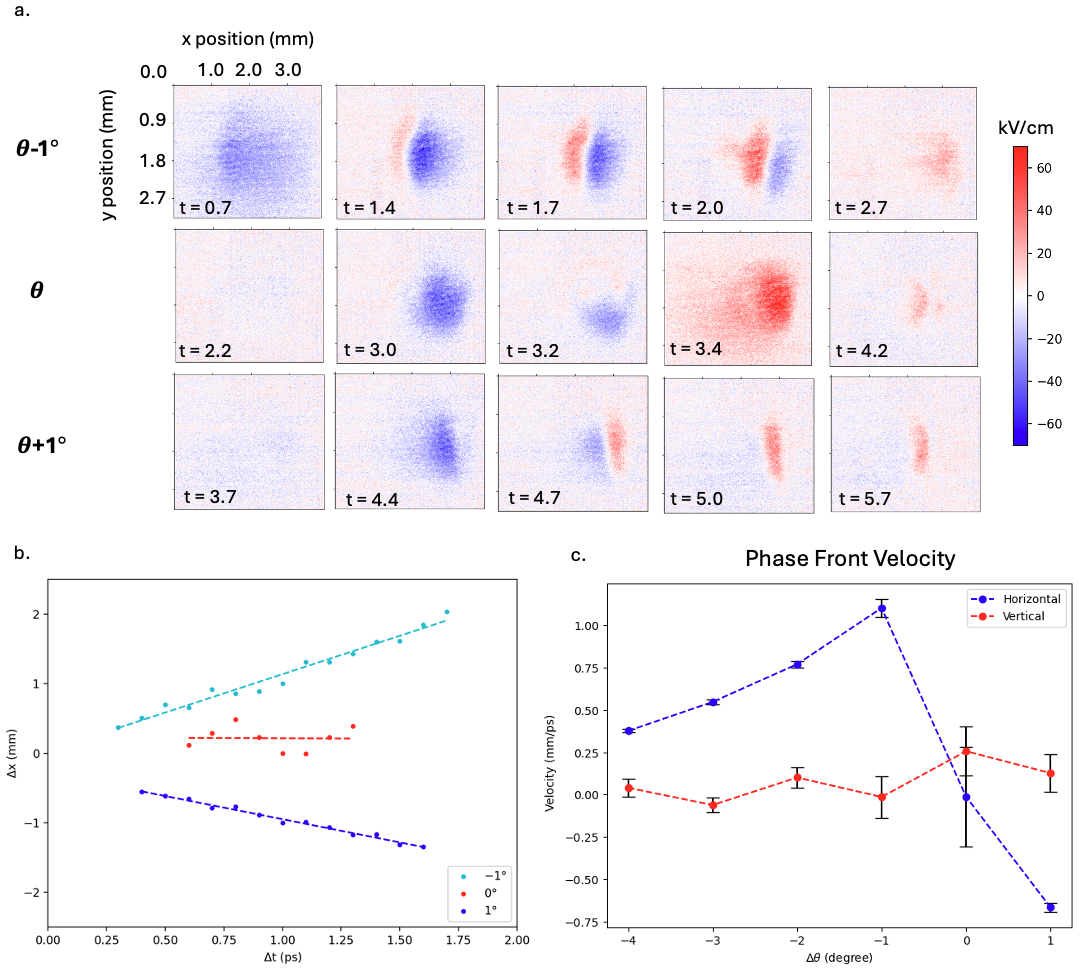}
\caption{Result of misalignment of the diffraction grating in the THz generation setup where (a) shows the THz field images for ideal alignment ($\theta$, as shown in Fig.1) and $\pm1\degree$ from ideal. A linear fit of the location of the peak field for each image was applied to every diffraction grating angle scan. Representative fits with time ranges and x position shifted for ease of viewing, are shown in (b). From these fits the vertical and horizontal velocity of the phase front can be calculated for each diffraction grating angle studied, as shown in (c). }
\label{diffraction_tilt_data}
\end{figure}

Slight adjustments in the diffraction grating angle (Fig. \ref{diffraction_tilt_diagram}) induced an angle in emission of the THz pulse from the LN surface (Fig. \ref{diffraction_tilt_data}). Plots of the peak field value for each image centered around the time of the absolute peak field allowed a linear fit to be applied to the data (Fig. \ref{diffraction_tilt_data} b.). This revealed the phase velocity of the field as it was emitted from the crystal and allowed the magnitude and direction to be quantified (Fig. \ref{diffraction_tilt_data} c.). The lateral and vertical phase velocity for the optimum case is close to 0 mm/ps as expected from theory. This analysis demonstrates that it is possible to tilt the phase front and induce a negative or positive phase velocity in the horizontal direction, and confirms that slight changes in the diffraction grating angle have a large effect on the spatiotemporal profile of the THz pulse.

\begin{figure}[htbp] 
\centering\includegraphics[scale=0.3]{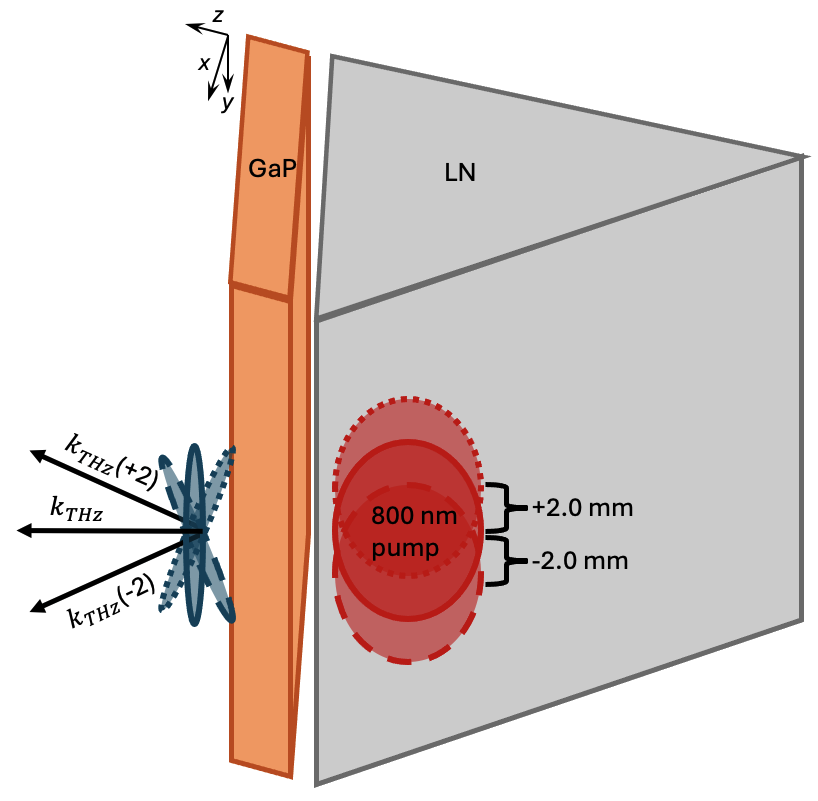}
\caption{Diagram showing the effect of slight vertical misalignment of the 800 nm pump on the generated THz field where $k_{THz}$ is ideal THz generation and $k_{THz}(\pm2)$ is slight misalignment.}
\label{vertical_tilt_diagram}
\end{figure}

\begin{figure}[htbp] 
\centering\includegraphics[scale=0.35]{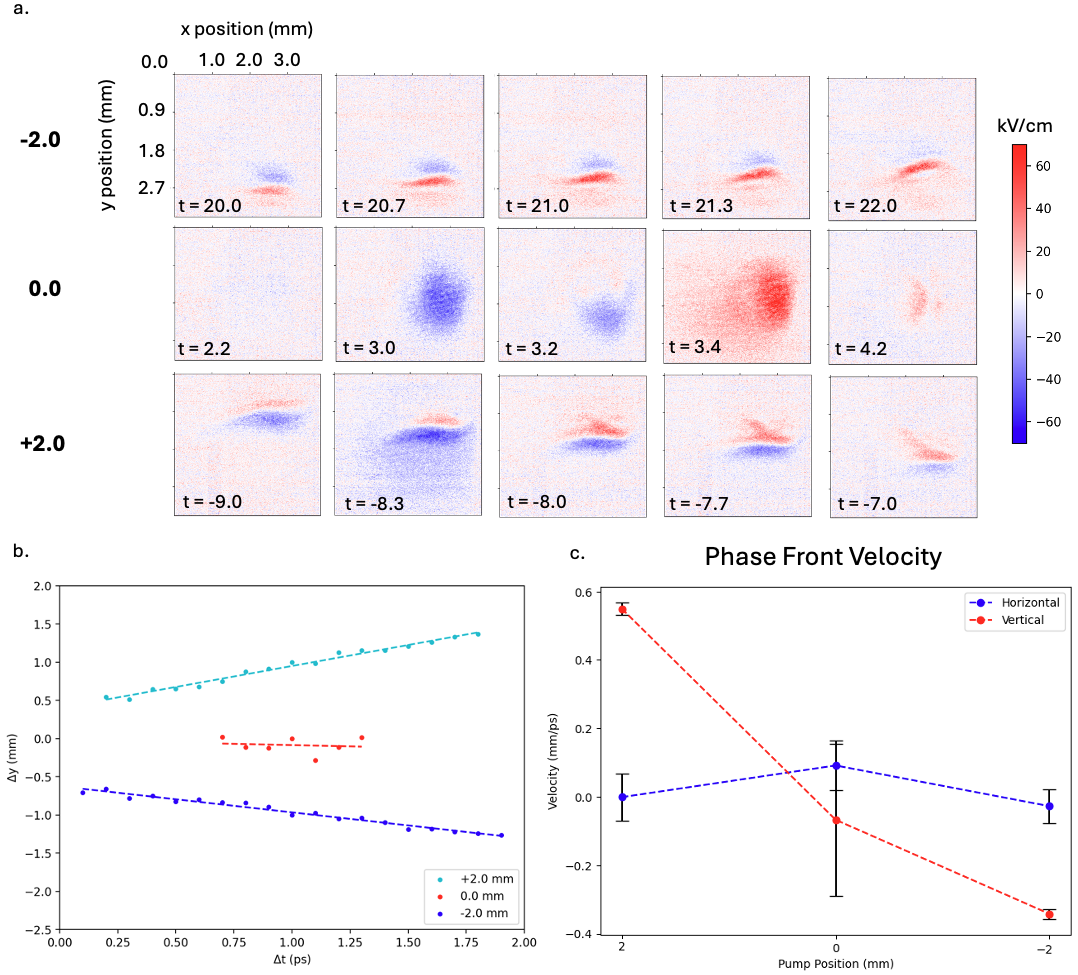}
\caption{Result of misalignment of the vertical position of the 800 nm pump on the input face of the LN crystal. (a) shows the THz field images for ideal alignment and $\pm2$mm vertical displacement. A linear fit of the location of the peak field for each image was applied to every vertical pump position scan. Representative fits with time ranges and y position shifted for ease of viewing, are shown in (b). From these fits the vertical and horizontal phase velocity can be calculated for each pump position, as shown in (c).}
\label{vertical_tilt_data}
\end{figure}

A similar effect was produced in the vertical direction by adjusting the vertical pointing of the final mirror before the telescope in the tilted pulse front generation setup (Fig. \ref{vertical_tilt_diagram}, \ref{vertical_tilt_data}). Using this mirror to adjust the 800 nm pump position on the LN input face by $\pm 2$mm induced a vertical phase velocity that can be analyzed using the same methods as the horizontal sweep (Fig. \ref{vertical_tilt_data} b, c.). This causes the bottom of the pulse to appear first for an upward tilt, and the reverse for a downward tilt (Fig \ref{vertical_tilt_data}a). These results show that the vertical pump position has a significant effect on the THz field and can be used to tilt the phase front and produce a negative or positive phase velocity in the vertical direction.

For both the vertical and horizontal sweep studies the observed features were consistent over multiple scans and re-optimization of the imaging and generation setup. Between each scan no parts of the experimental apparatus were adjusted except for the diffraction grating for the horizontal case, and the final mirror tilt for the vertical case. These minor adjustment of the diffraction grating angle ($\pm1\degree$) and final mirror tilt ($\pm2$ mm pump position) did not have significant effects on the peak field strength or frequency spectrum.

\subsection{Diffraction grating tilt}

\FloatBarrier

\FloatBarrier

\subsection{Final mirror vertical tilt}

\FloatBarrier

\FloatBarrier

\section{Discussion}

Using our 3D imaging technique, we completed a full characterization of the THz near-field at the LN crystal exit face, revealing features that were not possible to capture previously. To study these effects, we conducted a series of scans tilting the LN crystal horizontally and vertically, lowering pump power, varying LN-GaP distance, switching between a spherical and cylindrical telescope, adjusting the diffraction grating angle tilt, and adjusting the vertical pump position. This optimization revealed that varying the tilt of the diffraction grating angle and the incidence angle in the vertical direction had the most significant effect on the spatiotemporal properties of the THz near-field. In order to produce a spatially Gaussian THz pulse with a flat pulse front it is crucial to carefully tune these parameters. We also found that it was much simpler to produce a distortion-free pulse with spherical lenses in the telescope of the THz generation setup. When a telescope consisting of cylindrical lenses was used, small offsets in the rotation of the lenses produced an elliptical THz beam with strong phase front tilt. However, the effect induced by the cylindrical lenses was not as significant as that of the diffraction grating and final mirror tilt. All other parameters studied produced little to no change in the THz field profile besides reduction in overall intensity.

The temporal delay and spatial profile of the THz near-field will affect the out-coupling from LN and the choice of transport optics. Images of the THz near-field allows the design of specialized transport optics that more effectively capture and focus the THz pulse. In addition to being an effective method for studying the THz near-field, this method also functions as a real-time diagnostic for THz optimization. Using our time-dependent THz near-field imaging setup in the laboratory  enables real-time tuning and optimization of both the temporal delay and the spatial profile of the THz source. This facilitates the generation of THz fields with parameters tailored for specific applications. For instance, adjusting the temporal delay can be used to maximize energy transfer in THz-driven particle acceleration experiments.

\section{Conclusion}
In conclusion, we have successfully demonstrated a novel technique for quantitative spatiotemporal characterization of strong-field, single-cycle THz pulses generated via optical rectification in LiNbO$_3$. This method utilizes 2-D near-field electro-optic imaging, providing a complete 3D reconstruction of the THz near-field at the output face of the LN crystal. Our approach offers significant advantages over existing far-field methods and standard electro-optic sampling techniques, including enhanced spatial resolution and rapid data acquisition. We were able to reconstruct the full temporal evolution of the THz field and quantitatively analyze the impact of key experimental parameters on its spatiotemporal characteristics.

We found that careful adjustment of the diffraction grating angle and the vertical position of the pump beam significantly influence the phase front and the overall profile of the generated THz pulse. Specifically, precise control over these parameters is essential for achieving a spatially Gaussian THz pulse with a flat pulse front, crucial for optimizing THz generation, transport, and application. The ability to readily visualize and characterize the THz near-field allows for real-time optimization of THz generation parameters, paving the way for the design of customized transport optics and the tailoring of THz pulse characteristics for specific applications. For example, manipulating the temporal delay allows us to optimize energy transfer in THz-driven particle acceleration experiments.

This work represents a significant step forward in the characterization of high-field THz sources. The ability to visualize and manipulate the near-field properties opens new avenues for advancing THz science and its applications. Future work will focus on optimizing THz coupling and transport, improving the conversion efficiency, and extending the technique to explore the effects of different LN crystal parameters. The insights gained from this near-field imaging technique will contribute to a deeper understanding of the THz generation mechanism and facilitate the development of advanced THz technologies.
\begin{backmatter}
\bmsection{Funding}
Use of the Linac Coherent Light Source (LCLS), SLAC National Accelerator Laboratory, is supported by the U.S. Department of Energy, Office of Science, Office of Basic Energy Sciences under Contract No. DE-AC02-76SF00515

\bmsection{Disclosures}
The authors declare no conflicts of interest.

\bmsection{Data availability} Data underlying the results presented in this paper are not publicly available at this time but may be obtained from the authors upon reasonable request.\\
\end{backmatter}

\noindent See Supplement 1 for supporting content.


\bibliography{sample}

\noindent
\textbf{Supplemental document}

\begin{figure}[htbp]
    \centering
    \includegraphics[width=0.5\linewidth]{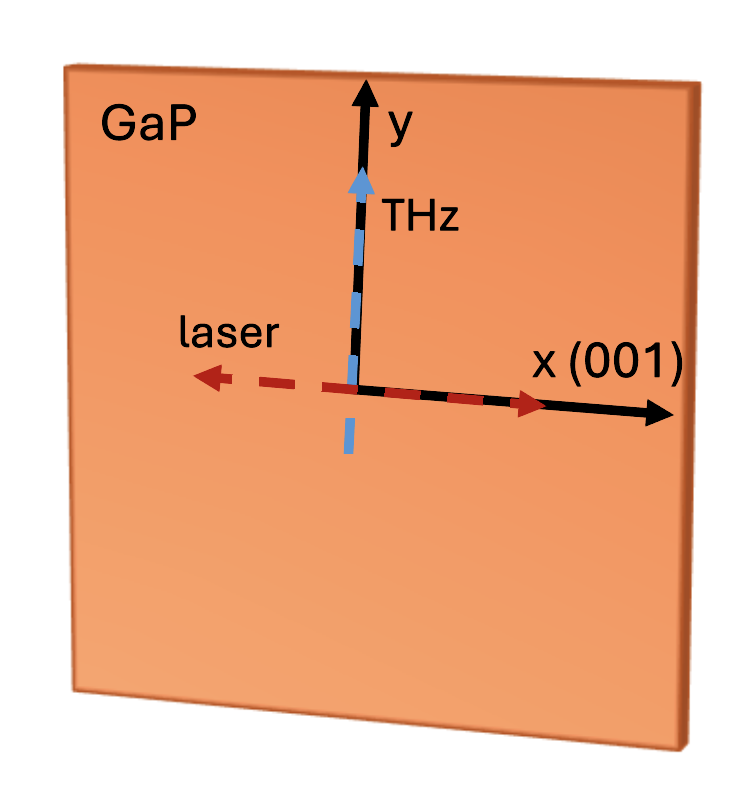}
    \caption{Diagram of the lab coordinate system used in the calculation with respect to GaP. The z axis is the reflected probe beam direction of travel (out of page). The dashed red and blue lines indicate the input probe laser polarization and the THz polarization respectively.} 
    \label{fig:EOSGaPcoords}
    \end{figure}

The method used to measure the THz near-field in this work is different than the standard EOS approach typically used to measure THz \cite{Planken:01}. Most significantly, in this method only the vertical component of the tilted probe beam is imaged. This means it is impossible to measure the relative difference between the vertical and horizontal polarization components of the probe beam, as in standard EOS. However, it is still possible to extract a THz field strength from the probe intensity measurements collected in our method by comparing the measured intensities on the camera with and without THz. To do this requires reworking the typical EOS theory from the Jones matrices of the optical components within the setup. This analytical calculation was completed, propagating the probe beam through the optics of the entire setup and solving for a relation between the THz field strength and the measured laser intensity. We begin with the definitions of the relevant Jones matrices for the optical components and the coordinate system used in the calculation (Fig. \ref{fig:EOSGaPcoords}). In the following, the subscript $i$ indicates the incident probe beam and the subscript $r$ indicates reflected, after reflection off the HR coating on the GaP crystal.
    \FloatBarrier
The first component is the polarizing beam splitter cube:
    \begin{gather}
    [PBS]_i = 
    \begin{bmatrix}
    1 & 0 \\
    0 & 0 
    \end{bmatrix}
    \end{gather}
    \begin{gather}
    [PBS]_r = 
    \begin{bmatrix}
    0 & 0 \\
    0 & 1 \\
    \end{bmatrix}
    \end{gather}
followed by a quarter waveplate at an arbitrary angle: 
    \begin{gather}
    [\lambda/4]_i =
    \mathrm{e}^{-i \frac{\pi}{4}}
    \begin{bmatrix}
    \cos^2(\gamma) + i\sin^2(\gamma) & (1-i)\sin(\gamma)\cos(\gamma) \\
    (1-i)\sin(\gamma)\cos(\gamma) & \sin^2(\gamma) + i\cos^2(\gamma) 
    \end{bmatrix}
    \end{gather}
    \begin{gather}
    [\lambda/4]_r = 
    \mathrm{e}^{-i \frac{\pi}{4}}
    \begin{bmatrix}
    \cos^2(-\gamma) + i\sin^2(-\gamma) & (1-i)\sin(-\gamma)\cos(-\gamma) \\
    (1-i)\sin(-\gamma)\cos(-\gamma) & \sin^2(-\gamma) + i\cos^2(-\gamma) 
    \end{bmatrix}
    \end{gather}
where $\gamma$ is the angle of the quarter waveplate fast axis with respect to the x axis. All Jones matrices for the optical components have been written in terms of the lab coordinate system (Fig. \ref{fig:EOSGaPcoords}), but the matrices for the GaP crystal must be written in the coordinate system of the refractive index ellipsoid within the crystal. Because of this we include a coordinate transfer matrix for a rotation about the z axis to convert from the lab coordinates to the refractive index ellipsoid coordinates:
    \begin{gather}
    [R(\theta)]_z = 
    \begin{bmatrix}
    \cos(\theta) & -\sin(\theta) \\
    \sin(\theta) & \cos(\theta)
    \end{bmatrix}
    \end{gather}
where $\theta$ is the angle over which the coordinate system is rotated and falls under the following equations \cite{Planken:01}
    \begin{equation} \label{eq:theta}
    2\theta = -\arctan(2\tan\alpha) - n\pi
    \end{equation}
    \begin{equation} \label{eq:alpha}
    \left( m - \frac{1}{2}\right)\pi \leq \alpha < \left(m + \frac{1}{2} \right)\pi, ~~ m = 0,1,......
    \end{equation}
where $\alpha$ is the angle between the THz field polarization and the GaP (001) axis. For this experiment, we are measuring a vertically polarized THz field, therefore $\alpha = \pi/2$, which yields $m=1$ and $\theta=-3\pi/4$. After rotation into the crystal coordinate system we can apply the Jones matrices for the GaP crystal \cite{YarivYeh}:
    \begin{gather}
    [GaP]_i=
    \begin{bmatrix}
    1 & 0 \\
    0 & 1 \\
    \end{bmatrix}
    \end{gather}
    \begin{gather}
    [GaP]_r = 
    \begin{bmatrix}
    \mathrm{e}^{-in_x\frac{\omega}{c}l} & 0 \\
    0 & \mathrm{e}^{-in_y\frac{\omega}{c}l} \\
    \end{bmatrix}
    \end{gather}
the incident GaP matrix is the identity matrix because the effect of the counter propagating THz and laser probe is negligible, so only the interaction after reflection off the back surface of the GaP ($[GaP]_r$) is significant. Here, $n_x,\ n_y$ are defined using the following equations \cite{Planken:01}:
    \begin{equation}
    n_{x}(\alpha) \approx n + \frac{n^3}{2}E_{THz}r_{41}[\cos\alpha \sin^2\theta + \cos(\alpha + 2\theta)]
    \end{equation}
    \begin{equation}
    n_{y}(\alpha) \approx n + \frac{n^3}{2}E_{THz}r_{41}[\cos\alpha \cos^2\theta - \cos(\alpha + 2\theta)]
    \end{equation}
where $n$ is the unperturbed refractive index for GaP, $\alpha$ is defined in Equation \ref{eq:alpha}, and $r_{41}$ is the electro-optic coefficient for GaP. Finally, the last matrix needed is for a mirror, to represent the probe reflection of the back surface of the GaP:
    \begin{gather}
    [mir] = 
    \begin{bmatrix}
    1 & 0 \\
    0 & -1 
    \end{bmatrix}
    \end{gather}
using these matrices we can now write out the equation for our electro-optic sampling setup:
    \begin{gather} \label{eq:cubeEOS}
    [PBS]_r[\lambda/4]_r[R(-\theta)]_z[GaP]_r[R(\theta)]_z[mir][GaP]_i[\lambda/4]_i[PBS]_i
    \begin{bmatrix}
    V_x \\
    V_y 
    \end{bmatrix} 
    = 
    \begin{bmatrix}
    V_x' \\
    V_y' 
    \end{bmatrix} 
    \end{gather}
where $V_x,\ V_y$ are the horizontal and vertical component of the input polarization vector of the laser and $V_x',\ V_y'$ are the respective output vector. Using this equation it is possible to calculate a relation between the measured camera intensity and the THz field strength. Assuming that the input laser is perfectly horizontally polarized ($V_y = 0$), the (001) axis of the GaP is parallel to the table, and the THz field is vertically polarized (perpendicular to the (001) axis), you can solve Equation \ref{eq:cubeEOS} for the laser intensity in the THz ON ($I_{THz}$) and THz OFF ($I_{B,\gamma}$) case:
    \begin{equation} \label{eq:ITHz_I0}
    I_{THz} = |V_{y}'(E_{THz})|^2,\ I_{B,\gamma} = |V_y'(E_{THz} = 0)|^2 
    \end{equation}
where $I_{THz},\ I_{B,\gamma} $ are the laser intensities measured on the camera. These two vaues can be combined to yield a relation between THz field strength and the laser intensity measured on camera after background subtraction and normalization ($I_c$):
    \begin{equation} \label{eq:I_c}
    I_c = \frac{I_{B,\gamma} - I_{THz}}{I_{B,\gamma}}
    \end{equation}
conversion from field strength inside the GaP crystal ($E_{GaP}$) to field strength in air ($E_{air}$) can be done using:
    \begin{equation} \label{eq:Eair}
    E_{air} = E_{GaP}/t_{EO},\ t_{EO} = \frac{2}{n_{THz}+1}
    \end{equation}
where $n_{THz}$ is the refractive index of GaP for THz. Applying known values for GaP and a range of values for $E_{THz}$ produces the plot in Figure 3 of the main text. All calculations (\ref{eq:cubeEOS} - \ref{eq:Eair}) were carried out in Python.

\end{document}